\documentclass[iop, twocolumn]{aastex631}
\hypersetup{citecolor=blue}
\usepackage{multirow}
\usepackage{amsmath}
\usepackage{graphicx}
\usepackage{subfigure}
\usepackage{hyperref}
\usepackage{nameref}
\usepackage{threeparttable}
\usepackage{tikz}
\usepackage{orcidlink}
\usetikzlibrary{calc, arrows.meta, positioning}

\newcommand{\df}{\mathrm{DF}}

\begin{document}

\title{Formation of Merging Black Hole Binaries Inside Massive AGN Stars}

\correspondingauthor{Sizheng Ma}
\email{sma2@perimeterinstitute.ca}

\newcommand{\westlake}{Department of Astronomy, Westlake University, Hangzhou, Zhejiang, 310030, China}

\newcommand{\tsinghuaias}{Institute for Advanced Study, 
Tsinghua University, Beijing, 100084, China}

\newcommand{\ucsc}{Department of Astronomy and Astrophysics, University of California, Santa Cruz, CA, 95064, USA}

\author{Sizheng Ma\,\orcidlink{0000-0002-4645-453X}}
\affiliation{Perimeter Institute for Theoretical Physics, Waterloo, ON N2L2Y5, Canada}

\author{Douglas N. C. Lin\,\orcidlink{0000-0001-5466-4628}}
\affiliation{\ucsc}

\author{David A. Velasco-Romero}
\affiliation{School of Natural Sciences, Institute for Advanced Study, NJ 08540, USA}
\affiliation{Department of Astrophysical Sciences, Princeton University, NJ 08544, USA}

\begin{abstract}
Stars embedded in the disks of active galactic nuclei (AGN) can grow to hundreds
of solar masses, and the same disks are expected to host a population of
stellar-mass black holes. A star may therefore capture passing black holes,
turning its interior into a potential factory for forming compact binary black
hole systems and, ultimately, for the gravitational-wave events seen by our detectors. We test how readily this channel operates using
three-dimensional hydrodynamic simulations of black hole--star encounters. As a proof of principle, we adopt a
star-to-black-hole mass ratio of $34\!:\!1$, and find that the capture does not
disrupt the structure of the star: it heats the star by $10$--$30\%$, and the
star loses only $\lesssim1\%$ of its mass throughout. The captured black hole loses its orbital angular momentum rapidly to dynamical friction,
sinking to the stellar center within a stellar dynamical time
$\lesssim10^{4}\,$s. If the star already harbors a black hole at its center, the
two form a bound binary whose gravitational-wave coalescence time falls below
$10^{4}\,$yr. Depending on the geometry and speed of the encounter, the resulting
orbit may be nearly circular or retain substantial eccentricity. Our calculations
confirm that black hole--star encounters in AGN disks are indeed a viable channel
for assembling stellar-mass black hole binaries and supplying sources for gravitational-wave detectors.
\end{abstract}


\section{Introduction}\label{sec:intro}

The rapidly growing catalog of compact-binary mergers observed by the LIGO--Virgo--KAGRA network has transformed gravitational-wave astronomy from the study of individual events into a powerful probe of black-hole populations and their astrophysical origins \citep{LVK2019Population,LVK2021GWTC2Population,LVK2023GWTC3Population,LVK2026GWTC5Population}. By measuring the distributions of their masses, spins, and orbital properties, population studies have begun to constrain the nature of their progenitor systems and the astrophysical environments in which these binaries form and evolve \citep[e.g.,][]{Gerosa:2017kvu,TalbotThrane2017,FishbachFarrHolz2020,CallisterFarr2024,GerosaFishbach2021,Mapelli2021,MandelFarmer2022,Tong:2025wpz,Banagiri:2025dmy,Plunkett:2026pxt}.

A broad range of astrophysical pathways have been proposed for the formation of merging binary black holes. Many of these operate in gas-poor, or ``dry'' environments. In isolated binaries, mass transfer and common-envelope evolution can shrink the orbit sufficiently for them to merge within a Hubble time \citep[e.g.,][]{2000ARA&A..38..113T,2013A&ARv..21...59I,Roepke:2022icg}. 
Dynamical encounters in dense stellar systems can likewise assemble and harden binaries \citep{PortegiesZwart:1999nm,2006ApJ...637..937O,2015ApJ...800....9M,Rodriguez2016,AntoniniRasio2016}. 
Alternatively, secular interactions in hierarchical stellar systems provide another pathway, driving otherwise wide binaries to large eccentricities and, in some cases, eventual merger \citep[e.g.,][]{SilsbeeTremaine2017,AntoniniToonenHamers2017}.

In contrast to these gas-poor channels, active galactic nuclei (AGN) disks provide a gas-rich, ``wet'' environment in which the surrounding gas can directly influence black-hole orbits. Here, interactions with the gas can dissipate black holes' orbital energy and angular momentum, promoting the capture of black
holes into the disk, their subsequent migration, and the formation and hardening
of binaries that ultimately merge through gravitational-wave emission \citep{Bartos2017,StoneMetzgerHaiman2017,McKernan2018,Bellovary2016,Secunda2019,Tagawa2020,Li2022,Li2023,Rowan2023,Baruteau2011,LiLai2022,Yang2026,Antoni2019,Tagawa2021a,Tagawa2021b,McKernan2020a,McKernan2020b}. 

Direct gas-driven orbital evolution, however, is not the only way in which AGN disks may facilitate the formation of compact binaries. It has been shown that sufficiently massive disks can become gravitationally unstable and fragment into massive stars \citep{Goodman2003,GoodmanTan2004,Jermyn2022,ChenLin2024,Cantiello2021,Fryer2025,Xu2026}. Reaching several hundred solar masses, such stars may be capable of capturing stellar-mass black holes without being disrupted \citep{Shi:2026kci}, thereby forming so-called ``black-hole stars'' or quasistars \citep{BegelmanRossiArmitage,VolonteriBegelman_2010,coughlin2024}, analogous in some respects to Thorne--Zytkow objects \citep{ThorneZytkow_1977}. If such captures occur repeatedly, a single massive star may host multiple black holes, raising the possibility that the captured black holes could be assembled into compact binaries that later become gravitational-wave sources. Very recently, this possibility has begun to be explored in detail. Using semi-analytic modeling, \citet{Cantiello2026} showed that this channel can produce binaries detectable by the LIGO--Virgo--KAGRA network and estimated a merger rate up to $\sim 8\,{\rm Gpc}^{-3}\,{\rm yr}^{-1}$. Complementarily, \citet{Hu:2026ica} performed hydrodynamical simulations with the Lagrangian meshless code \textsc{GIZMO} \citep{Hopkins:2014qka}, demonstrating that the channel can yield highly eccentric compact binaries.

In this work, we provide an independent hydrodynamic investigation of this proposed formation channel using a grid-based numerical approach. Using a distinct numerical method, our simulations confirm that massive AGN stars are able to assemble compact binaries. 
We explore a region of parameter space distinct from that considered by \citet{Hu:2026ica}, including a lower black-hole-to-star mass ratio, a different equation of state, and different encounter kinematics, and find that the resulting binaries need not be exclusively highly eccentric. We further examine the hydrodynamic response of the host star, including the formation of gravitational wakes, the excitation of acoustic disturbances, and the associated stellar heating and mass loss. Our results therefore complement those of \citet{Hu:2026ica}.

The rest of this paper is organized as follows. In Sec.~\ref{sec:Setup}, we describe the numerical setup and the star--black-hole encounter configurations considered in our simulations. In Sec.~\ref{sec:results}, we present the resulting orbital evolution and hydrodynamic response of the star. Finally, we summarize our main findings and discuss their implications and future directions in Sec.~\ref{sec:conclusion}.

\section{Simulation Setup}\label{sec:Setup}
We evolve the star with the grid-based code \textsc{AthenaK} \citep{Stone2024AthenaK}, solving ideal, inviscid, self-gravitating hydrodynamics on a three-dimensional Cartesian mesh. The continuity, momentum, and total-energy equations are
\begin{subequations}
\label{eq:eom_hydro}
    \begin{align}
\frac{\partial\rho}{\partial t}+\nabla\!\cdot\!(\rho\mathbf v)
  &= \dot S_\rho, \label{eq:cont}\\[0.3ex]
\frac{\partial(\rho\mathbf v)}{\partial t}
  +\nabla\!\cdot\!\big(\rho\mathbf v\mathbf v+P\,\mathbb{I}\big)
  &= -\rho\nabla(\Phi+\Phi_{\rm BH})+\dot{\mathbf S}_{\rm BH}, \label{eq:mom}\\[0.3ex]
\frac{\partial E}{\partial t}
  +\nabla\!\cdot\!\big[(E+P)\mathbf v\big]
  &= -\rho\mathbf v\!\cdot\!\nabla(\Phi+\Phi_{\rm BH})+\dot S_E, \label{eq:ene}
\end{align}
\end{subequations}
Here $\rho$, $\mathbf v$, and $P$ denote the gas density, velocity, and pressure, respectively, while $E=\rho\,e+\tfrac12\rho|\mathbf v|^2$ is the total energy density and $e$ is the specific internal energy. We close the system with an ideal-gas equation of state,
\begin{equation}
P=(\gamma-1)\,\rho\,e,\qquad \gamma=\tfrac53 .
\label{eq:eos}
\end{equation}
The self-gravitational potential of the gas, $\Phi$, obeys
\begin{equation}
\nabla^2\Phi = 4\pi G\rho,
\label{eq:poisson}
\end{equation}
which is solved with a geometric multigrid method \citep{TomidaStone2023}.

The black holes are evolved as point particles. Their combined gravitational potential is
\begin{equation}
\Phi_{\rm BH}=\sum_i\phi_{i},
\end{equation}
with each contribution modeled by a Plummer-softened point-mass potential,
\begin{equation}
\phi_{i}(\mathbf r)
=-\frac{Gm_i}
{\left(|\mathbf r-\mathbf r_i|^2+\varepsilon_i^2\right)^{1/2}}.
\end{equation}
We tie the softening scale $\varepsilon_i$ to the local Bondi--Hoyle--Lyttleton radius \citep{BondiHoyle1944,Bondi1952},
\begin{align}
    \varepsilon_i=\frac{2G\,m_i}{c_s^2+v_{i}^2}, \label{eq:def_epsilon_i}
\end{align}
where $c_s$ is the local sound speed and $v_i$ is the velocity of black hole $i$. We cap the softening length at $\varepsilon_i<0.1R_\star$ so that it does not grow to a substantial fraction of the star.

Each black hole feels the gravitational field of both the gas and its companion. Its equation of motion is
\begin{equation}
\frac{d^2\mathbf r_i}{dt^2}
=-\nabla\Phi_{pp,i}(\mathbf r_i)-\nabla\Phi(\mathbf r_i)
 \label{eq:BH_EOM}
\end{equation}
Here $\Phi_{pp,i}$ is the potential generated by the other black hole. We compute $\nabla\Phi$ on the mesh with second-order centered finite differences and interpolate it trilinearly to the particle position. Equation~\eqref{eq:BH_EOM} is integrated alongside the hydrodynamic system with a velocity-Verlet scheme.


\subsection{Sink terms}
\label{sec:sink}
The sink terms $\dot S_\rho$, $\dot{\mathbf S}_{\rm BH}$, and $\dot S_E$ in Eq.~\eqref{eq:eom_hydro} remove mass, momentum, and energy from gas near each black hole. We use the torque-free sink prescription given in \citep{2020ApJ...892L..29D,2021ApJ...921...71D},
\begin{subequations}
    \begin{align}
        &\dot S_\rho=-\rho\,\mathcal W(r),\\
& \dot{\mathbf S}_{\rm BH}=-\rho\,\mathcal W(r)\,\mathbf v^{\rm res},\\
&\dot S_E=-\rho\,\mathcal W(r)\,\mathcal E^{\rm res},
\label{eq:sinksrc}
    \end{align}
\end{subequations}
where 
\begin{align}
    \mathbf v^{\rm res}=\big[(\mathbf v-\mathbf v_{\rm BH})\! \cdot\!\hat{\mathbf r}\big]\,
  \hat{\mathbf r}+\mathbf v_{\rm BH}, \qquad 
\end{align}
and
\begin{equation}
\hat{\mathbf r}=
\frac{\mathbf r-\mathbf r_{\rm BH}}
{r}, \quad r=|\mathbf r-\mathbf r_{\rm BH}|.
\end{equation}
In the black-hole comoving frame, this construction removes only the radial component of the gas momentum, while leaving the tangential component untouched, in order to avoid a spurious torque. The associated specific energy removal is
\begin{equation}
\mathcal E^{\rm res}=\tfrac12\,|\mathbf v^{\rm res}|^2+e.
\end{equation}
We adopt the kernel
\begin{equation}
\mathcal W(r)=\frac{1}{t_s}\,\exp\!\left[-\left(\frac{r}{\varepsilon_i}\right)^{4}\right],
\label{eq:kernel}%
\end{equation}
where $t_s$ is tied to the Keplerian free-fall time at $\varepsilon_i$,
\begin{equation}
\frac{1}{t_s}=10^{-3}\,\sqrt{\frac{G\,m_i}{\varepsilon_i^{3}}}.
\label{eq:ts}
\end{equation}


Mass and linear momentum removed from the grid are deposited onto the black holes, so the combined fluid--particle system conserves both:
\begin{equation}
m_{\rm BH} \to m_{\rm BH}+\Delta M,\qquad
\mathbf{v}_{\rm BH} \to
\frac{m_{\rm BH}\,\mathbf{v}_{\rm BH}+\Delta\mathbf{P}}{m_{\rm BH}+\Delta M},
\label{eq:feedback}
\end{equation}
where $\Delta M=\int \Delta\rho\,dV$ and $\Delta\mathbf{P}=\int \Delta\rho\,\mathbf{v}^{\rm res}\,dV$, with the integrals taken over the sink cells.
Once the black holes approach closely enough that their separation, $d=|\mathbf r_1-\mathbf r_2|$, satisfies
\begin{equation}
d<\left(R_{{\rm BHL},1}+R_{{\rm BHL},2}\right),
\end{equation}
we replace the two individual sinks by a single sink centered on the binary center of mass. Its characteristic radius and removal time are again given by Eqs.~\eqref{eq:def_epsilon_i} and \eqref{eq:ts}, now using the total binary mass. Material removed by this merged sink is added to the binary, and the corresponding mass and momentum increments are divided between the two black holes in proportion to their masses.  

As shown below, although this prescription self-consistently accounts for the transfer of mass and momentum from the gas to the black holes, the resulting changes in their masses and momenta remain small in all of our simulations, which span only a few stellar dynamical times.



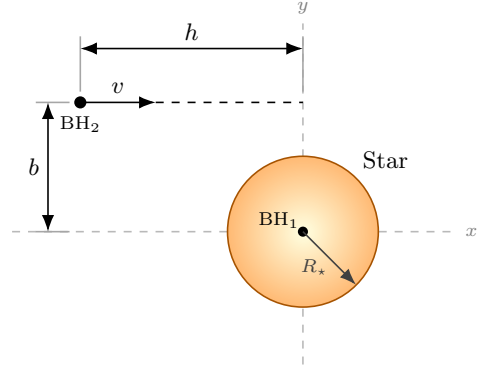
\begin{figure}[t]
\centering
\begin{tikzpicture}[scale=0.95, >={Latex[length=2.2mm]}, line width=0.6pt]
\draw[gray!55, dashed] (-4.05,0) -- (2.15,0);
\draw[gray!55, dashed] (0,-1.85) -- (0,2.9);
\node[gray, font=\scriptsize] at (2.35,0) {$x$};
\node[gray, font=\scriptsize] at (0,3.12) {$y$};
\shade[inner color=yellow!12, outer color=orange!55] (0,0) circle (1.05);
\draw[orange!65!black] (0,0) circle (1.05);
\node[font=\small] at (1.15,1.0) {Star}; \fill (0,0) circle (2pt); \node[font=\scriptsize, above left=-2pt and -2pt] at (0,0) {BH$_1$}; \draw[->, gray!50!black] (0,0) -- (0.74,-0.74) node[midway, below left=-4pt, font=\scriptsize] {$R_\star$};
\fill (-3.1,1.8) circle (2.4pt);
\node[font=\scriptsize, below=2pt] at (-3.1,1.8) {BH$_2$};
\draw[->] (-3.1,1.8) -- (-2.05,1.8) node[midway, above, font=\small] {$v$};
\draw[dashed] (-2.05,1.8) -- (0,1.8);
\draw[gray!70] (-3.1,1.95) -- (-3.1,2.72);
\draw[gray!70] (0,1.95) -- (0,2.72);
\draw[<->] (-3.1,2.55) -- (0,2.55) node[midway, above, font=\small] {$h$};
\draw[gray!70] (-3.72,1.8) -- (-3.25,1.8);
\draw[gray!70] (-3.72,0) -- (-3.25,0);
\draw[<->] (-3.55,1.8) -- (-3.55,0) node[midway, left, font=\small] {$b$};
\end{tikzpicture}
\caption{Schematic of the system considered in our simulations. The intruding black hole (BH$_2$) is initialized with impact parameter $b$, upstream offset $h$, and velocity $v$. The other black hole (BH$_1$) is initially located at the center of a star of radius $R_\star$.}
\label{fig:setup}
\end{figure}

\begin{figure*}[!htb]
    \includegraphics[width=0.98\textwidth]{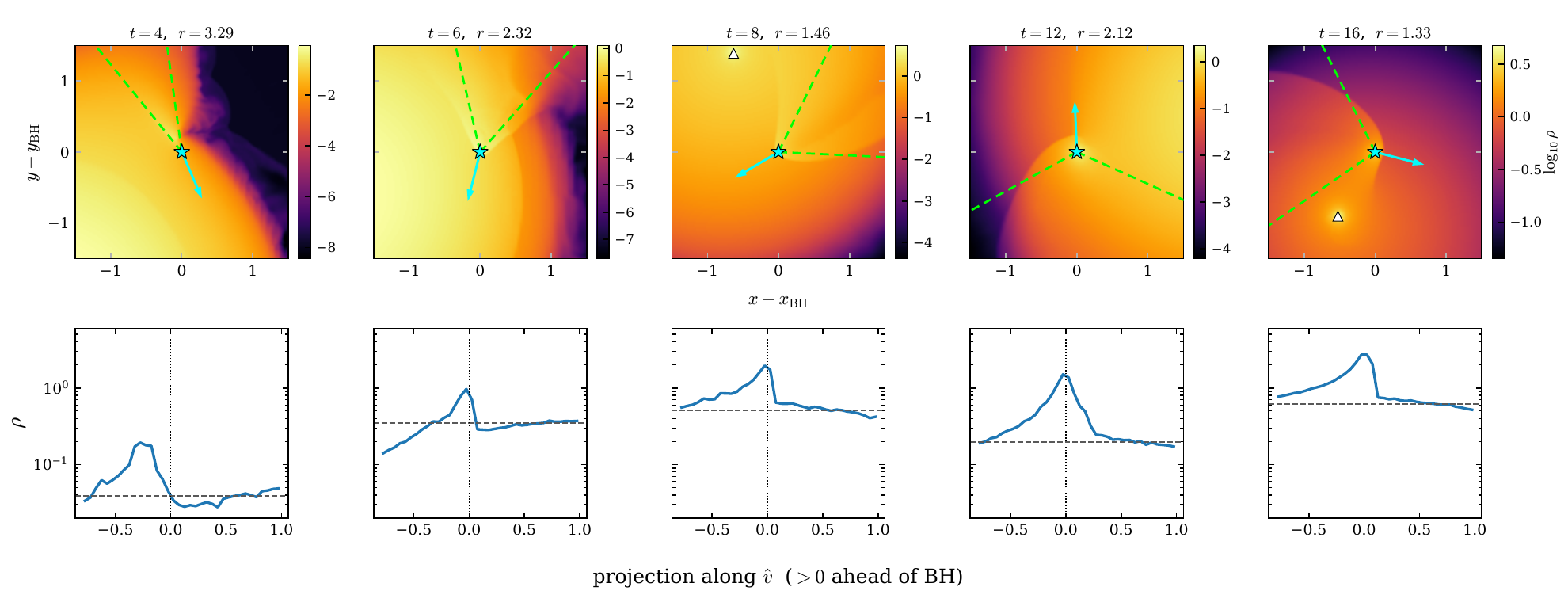}
    \caption{Gravitational wake produced by the intruding black hole during the early inspiral of run R1, shown at five representative times (columns). Top: Stellar density in the orbital plane, centered on the intruding black hole (cyan star). The cyan arrow indicates the instantaneous direction of motion, while the green dashed lines mark the linear Mach-cone half-opening angle, $\arcsin(1/\mathcal{M})$, evaluated using the local sound speed. Bottom: One-dimensional density profiles extracted along the instantaneous velocity direction, $\hat{v}$, with positive distance corresponding to the region ahead of the black hole; the vertical dotted line marks its position. The overdensity is systematically enhanced on the trailing side, producing the front--back asymmetry responsible for the dynamical friction that drives the orbital decay.}
    \label{fig:fig_wake_phase_inferno}
\end{figure*}
\subsection{Initial condition and evolution}
As shown in Fig.~\ref{fig:setup}, the star is initialized from a Lane--Emden solution with radius $R_\star=3.654$ and mass $M_\star=34.11$ in code units, with the first black hole initially at rest at the stellar center. We first relax the star--black-hole system toward hydrostatic equilibrium without the sink terms described in Sec.~\ref{sec:sink}. After this relaxation, the second black hole is introduced with impact parameter $b$, upstream offset $h$, and speed $v$. Table~\ref{tab:runs} lists four configurations considered in this work. For convenience, we quote $v$ relative to the Keplerian circular speed
\begin{align}
    v_c=\frac{(GM_\star)^{1/2}}{(b^2+h^2)^{1/4}},
\end{align}
evaluated at the initial separation. In all runs, the two black holes have equal masses $m_1=m_2=1$ (in code units).

We evolve the system in a Cartesian domain $[-20,20]^3$ with outflow boundaries. The root mesh contains $128^3$ cells, partitioned into mesh blocks of $16^3$ cells, giving a root-level spacing $\Delta x_{\rm root}=0.3125$.
We use block-based adaptive mesh refinement throughout. Two fixed nested regions, $[-10,10]^3$ and $[-5,5]^3$, are maintained at refinement levels 1 and 2 (the root grid is level 0). The immediate neighborhood of each black hole is refined more aggressively: a $[-0.15,0.15]^3$ box centered on each particle is covered by nested grids down to a finest spacing
\begin{equation}
\Delta x_{\min} =
\frac{\Delta x_{\rm root}}{2^{\,N_{\rm lev}-1}},
\end{equation}
with $N_{\rm lev}=8$ in the highest-resolution calculations.

The hydrodynamic update uses PPM4 reconstruction, the HLLE Riemann solver, and a second-order Runge--Kutta integrator. We set the CFL number to 0.4.


\begin{table}[htb]
\centering
\caption{Initial conditions for the four simulations. The columns give the impact parameter $b$, upstream distance $h$ from closest approach, and initial speed $v$ of BH$_2$ (see Figure~\ref{fig:setup}). 
In all models, $R_\star=3.654$, $M_\star=34.11$, and $m_1=m_2=1$.
All quantities are in code units.}
\begin{tabular}{lcccc}
\hline
Run & R1  & R2  & R3  & R4  \\
\hline
$b$ & 4.0 & 4.0 & 1.8 & 1.8  \\
$h$ & 0.0 & 0.0 & 4.0 & 4.0  \\
$v$ & $0.73v_c$ & $v_c$ & $v_c$ & $2v_c$  \\
\hline
\end{tabular}
\label{tab:runs}
\end{table}

\section{Results}\label{sec:results}
\subsection{R1}
We begin with run R1 in Table~\ref{tab:runs}. Here the exterior black hole starts at $(b,h)=(4.0,0)$ with $v=0.73\,v_c$. The top row of Figure~\ref{fig:fig_wake_phase_inferno} follows the stellar density in the orbital plane through five stages of the early inspiral. The black hole enters through the dilute outer envelope at supersonic speed, driving a bow shock and leaving an overdense gravitational wake behind it.

The green dashed lines indicate the linear Mach-cone half-opening angle, $\arcsin(1/\mathcal{M})$. Note that the expression is derived for a uniform medium, unlike the strongly stratified stellar envelope here, so we use it only as a geometric reference. Deeper inside the star, the local Mach number falls toward unity and the wake broadens accordingly.

\begin{figure}[!htb]
    \includegraphics[width=\columnwidth]{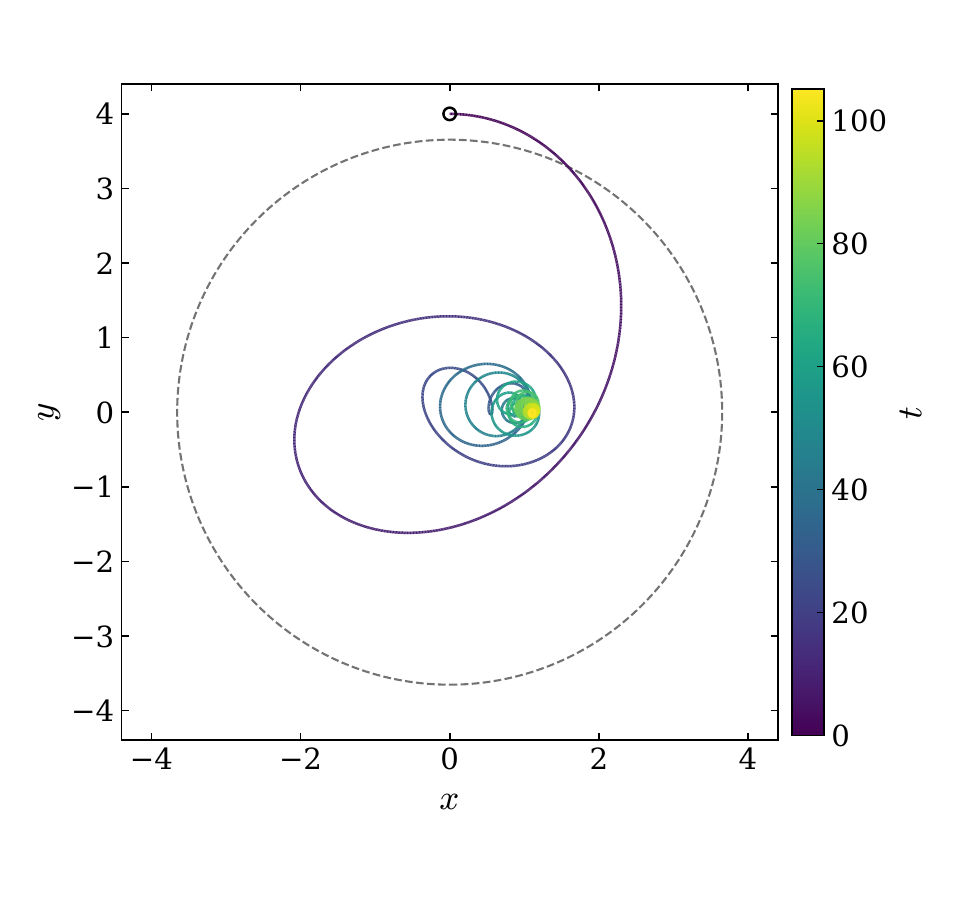}
    \caption{Orbital-plane trajectory of the intruding black hole in run R1, colored
    by time from release at $t=0$ to the end of the run at $t\simeq105$. The open
    circle marks the launch point (Table~\ref{tab:runs}), and the gray dashed circle
    marks the initial stellar surface $R_\star=3.654$. }
    \label{fig:fig_traj_res3}
\end{figure}

\begin{figure*}[!htb]
    \includegraphics[width=0.98\textwidth]{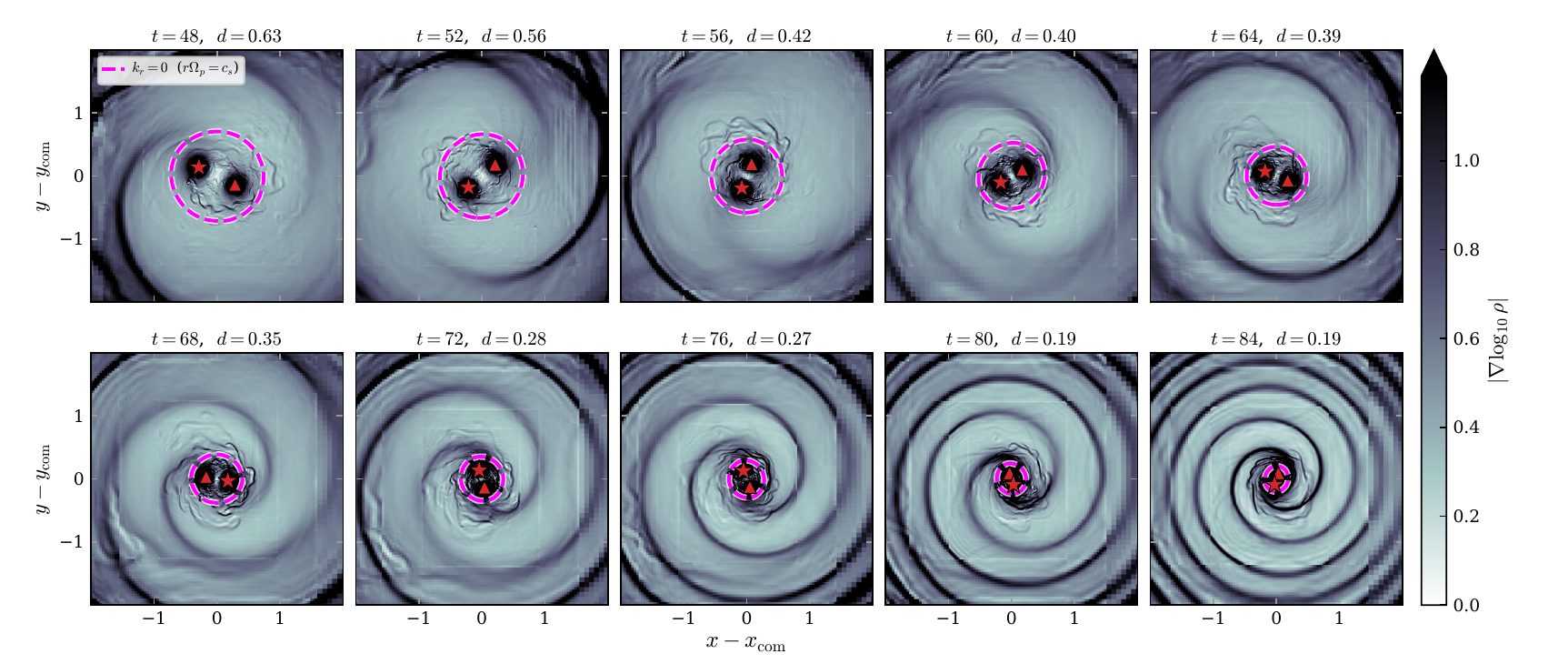}
    \caption{Two-armed acoustic spiral excited in the stellar interior during the late inspiral of run R1. The magnitude of the density gradient, $|\nabla\log_{10}\rho|$, in the orbital plane ($z=0$) is shown at several simulation times. The red star and triangle mark the positions of the two black holes. The magenta dashed circle marks the inner turning point of the $m=2$ response, defined by $k_r=0$ in Eq.~\eqref{eq:disp}. All quantities are given in code units.}
    \label{fig:spiral_late}
\end{figure*}

The bottom row of Figure~\ref{fig:fig_wake_phase_inferno} makes the asymmetry more explicit. We extract one-dimensional density profiles along the instantaneous direction of motion, $\hat v$, taking positive distance to lie ahead of the black hole. The excess density is weighted toward the trailing side. This front--back imbalance exerts a net gravitational drag, namely dynamical friction \citep{1943ApJ....97..255C,1999ApJ...513..252O,2007ApJ...665..432K,2001MNRAS.322...67S,Kim:2008ab,ONeill:2024tnl,2026arXiv260716422G}, and removes orbital energy rapidly.

The trajectory in Figure~\ref{fig:fig_traj_res3} shows the consequence. The solid curve follows the intruding black hole, and the dashed circle marks the initial stellar surface. Within $\mathcal{O}(10)$ orbital cycles, the intruding black hole sinks to the stellar center and becomes gravitationally bound to the other black hole on a dynamical timescale.

\begin{figure}[!htb]
    \includegraphics[width=\columnwidth]{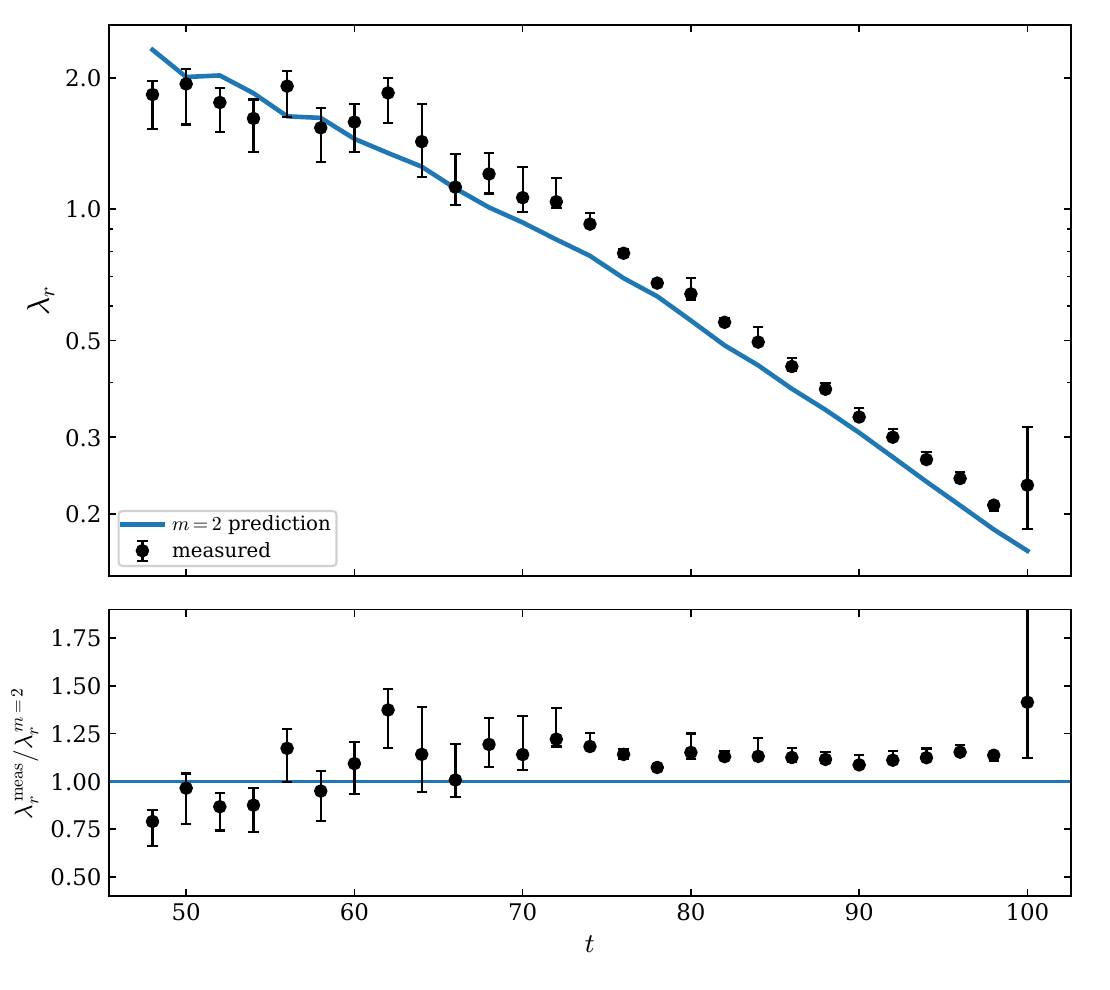}
    \caption{Radial wavelength of the spiral in run R1. Top:
    $\lambda_r$ measured from the simulation over $r\in[0.6,1.8]$ (points with
    error bars), compared with the linear $m=2$ prediction of
    Eq.~\eqref{eq:disp} evaluated at the instantaneous orbital frequency (blue
    line). Bottom: ratio of the measured to the predicted wavelength. The measurement tracks the prediction throughout the
    inspiral.}
    \label{fig:spiral_dispersion}
\end{figure}

During the subsequent late inspiral, the rotating $m=2$ component of the binary potential drives the stellar gas and produces a two-armed trailing spiral in the stellar interior, as shown in Fig.~\ref{fig:spiral_late}.
We interpret this morphology in terms of the linear acoustic response, whose radial wavenumber reads \citep{1980ApJS...43..469T}
\begin{equation}
k_r^2 \;=\; \left(\frac{m\,\Omega_{\rm orb}}{c_s}\right)^{2}-\left(\frac{m}{r}\right)^{2},
\qquad \lambda_r=\frac{2\pi}{k_r}.
\label{eq:disp}
\end{equation}
Here $\Omega_{\rm orb}$ is the orbital angular velocity of the binary. In Figure~\ref{fig:spiral_late}, the magenta dashed circles mark the $k_r^2=0$ boundary, separating the wave-like response from the near-zone regime. As the binary hardens and $\Omega_{\rm orb}$ increases, this boundary moves inward. 
Figure \ref{fig:spiral_dispersion} provides a quantitative check of this picture. We measure the radially averaged spiral wavelength, $\lambda_r$, over $r\in[0.6,1.8]$ and compare it with the prediction of Eq.~\eqref{eq:disp}. The measured wavelength decreases as the orbit shrinks, closely tracking the linear prediction throughout the evolution. A small systematic offset remains, plausibly reflecting the radial stratification of the stellar background and nonlinear coupling between modes.

\subsubsection{Dynamical friction}

\begin{figure}[!htb]
    \includegraphics[width=\columnwidth]{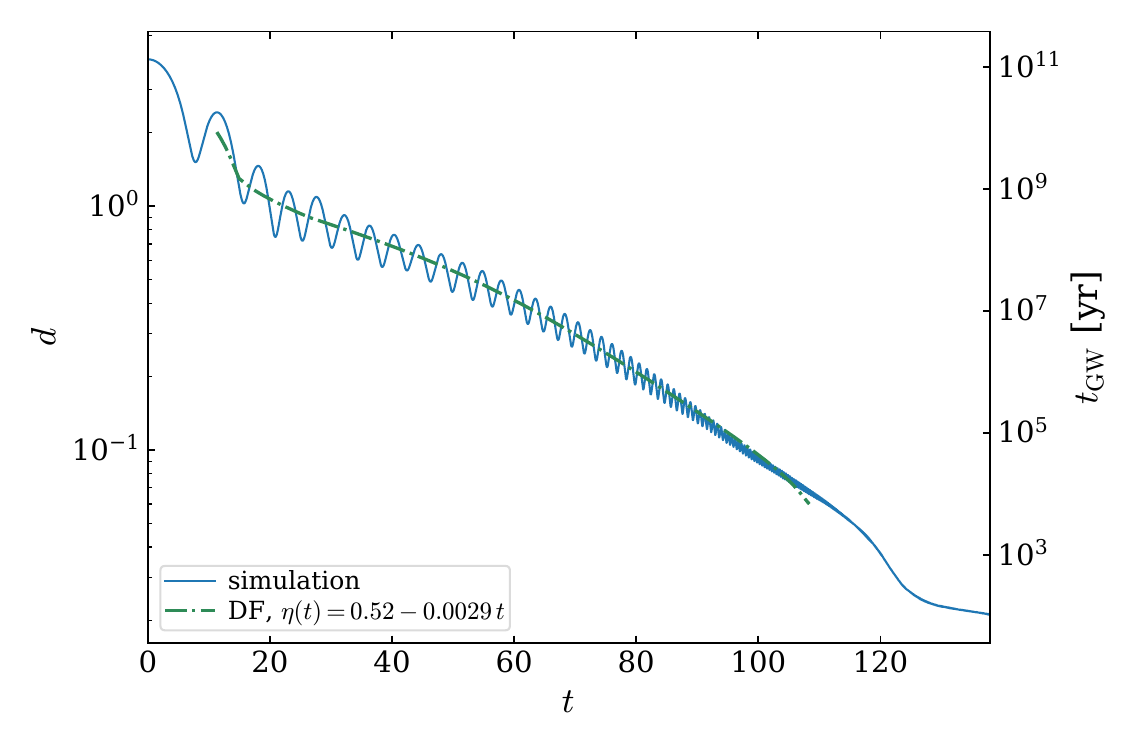}
    \caption{Orbital decay in run R1. The binary separation $d(t)$ measured in the simulation (blue, in code units) is compared with the orbit-averaged dynamical-friction model of Eq.~\eqref{eq:ostriker} (green dash-dashed). The orbital eccentricity decreases as the inspiral proceeds. The right axis shows the gravitational-wave coalescence time $t_{\rm GW}$ for a circular, equal-mass binary at the same separation, assuming $M_\star=172M_\odot$, $R_\star=20R_\odot$, and $m_1=m_2=5.0M_\odot$.}
    \label{fig:sep_res3}
\end{figure}

Throughout the simulation, the binary hardens rapidly, transferring orbital angular momentum to the stellar envelope through dynamical friction and the excitation of acoustic waves. Figure~\ref{fig:sep_res3} shows the evolution of the binary separation. 
The system is initially eccentric and progressively circularizes as the inspiral proceeds. At late times, the separation levels off at a scale comparable to the sink radius.

For comparison with the simulation, we model the gas drag using the dynamical-friction prescriptions of \cite{1999ApJ...513..252O,2007ApJ...665..432K,2001MNRAS.322...67S},
\begin{equation}
F_\df=\frac{4\pi\rho\,(Gm)^2}{v^2}\;I(\mathcal{M}),
\label{eq:ostriker}
\end{equation}
with 
\begin{equation}
    I=\begin{cases}
\displaystyle \eta(t) \left[\frac12\ln\!\frac{1+\mathcal{M}}{1-\mathcal{M}}-\mathcal{M}\right],
& \mathcal{M}<1 \ \ (\text{subsonic}),\\[2.2ex]
\displaystyle \frac12\ln\!\Bigl(1-\frac{1}{\mathcal{M}^2}\Bigr)+\ln \left(\frac{d}{2.25\varepsilon_i}\right),
& \mathcal{M}>1 \ \ (\text{supersonic}),
\end{cases}
\end{equation}
where $\varepsilon_i$ is given by Eq.~\eqref{eq:def_epsilon_i}. In the subsonic branch, we introduce a time-dependent factor $\eta(t)$ to account phenomenologically for deviations from the idealized assumptions underlying the standard dynamical-friction prescription, including the evolving orbit and the nonuniform, time-dependent stellar background.
The best-fitting form, $\eta(t)=0.52-0.0029\,t$, reproduces the orbit-averaged decay well (green dashed curve in Figure~\ref{fig:sep_res3}).

\begin{figure*}[!htb]
    \includegraphics[width=0.98\textwidth]{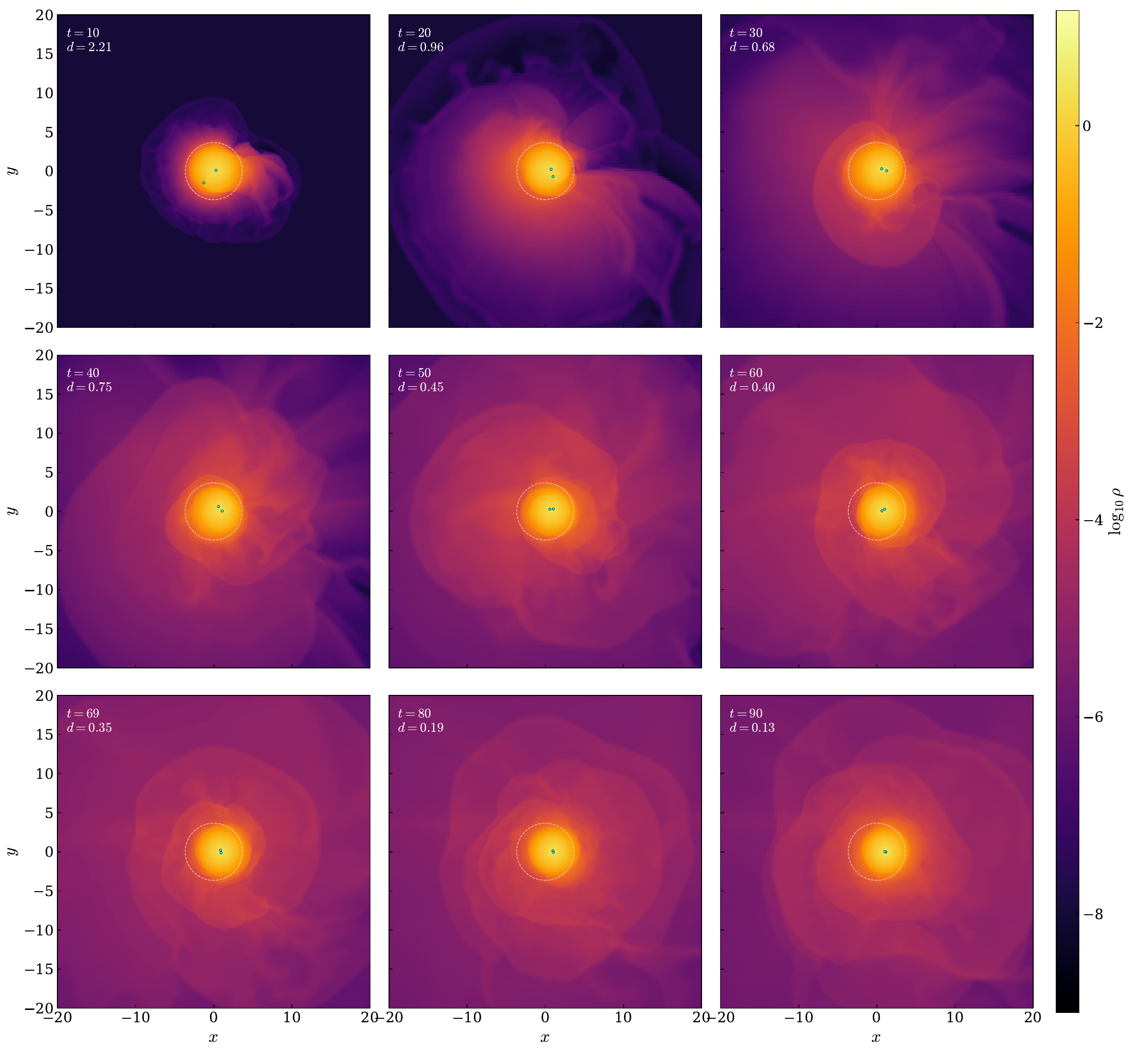}
    \caption{Global response of the star in run R1. Panels
    show the orbital-plane density $\log_{10}\rho$ over the full computational
    domain at nine times spanning $10\le t\le90$. The two small cyan circles mark the black holes,
    and the white dashed circle marks the initial stellar surface,
    $R_\star=3.654$. All quantities are in code units.}
    \label{fig:fig_dens_xy_domain_series_fullrange}
\end{figure*}

\subsubsection{Gravitational waves}
Gravitational radiation strengthens as the binary contracts, but it is still dynamically subdominant over the portion of the inspiral that we resolve. Before it can take over, the orbital separation approaches the sink scale and the calculation becomes increasingly sensitive to sink size and numerical resolution. We therefore do not evolve the system through a gravitational-wave-driven merger. Instead, we use the resolved separation to estimate the maximum remaining timescale, assuming that gravitational radiation alone governs the subsequent evolution.

For a circular binary, we have \citep{Peters:1963ux}
\begin{equation}
\begin{aligned}
t_{\rm GW} \simeq{}&
3.75\times10^4\ {\rm yr}\,
\left(\frac{d}{0.5R_\odot}\right)^4 \\
&\times
\left(\frac{m_1}{5M_\odot}\right)^{-1}
\left(\frac{m_2}{5M_\odot}\right)^{-1}
\left(\frac{m_1+m_2}{10M_\odot}\right)^{-1}.
\end{aligned}
\label{eq:tgw}
\end{equation}
Considering the physical scaling $M_\star=172M_\odot$ and $R_\star=20R_\odot$, the corresponding $t_{\rm GW}$ is shown on the right axis of Figure~\ref{fig:sep_res3}. By the end of the simulation, the inferred maximum remaining time has fallen to $\mathcal{O}(10^3)\,{\rm yr}$.

\begin{figure*}[!htb]
    \includegraphics[width=0.98\textwidth]{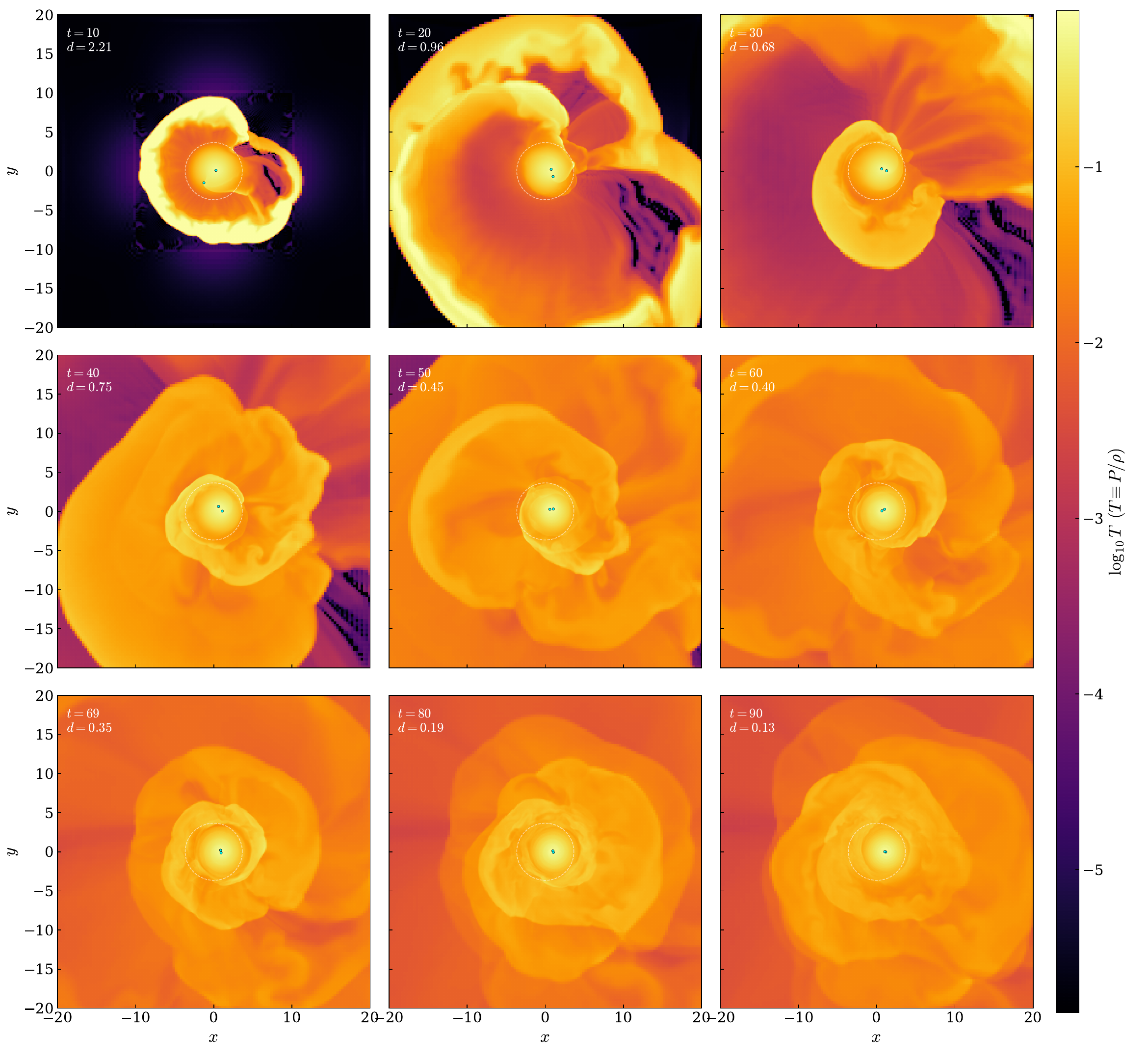}
    \caption{Similar to Fig.~\ref{fig:fig_dens_xy_domain_series_fullrange}, stellar temperature across the simulation domain in run R1.}
    \label{fig:fig_temp_xy_domain_series}
\end{figure*}


\subsubsection{End state of the star}
Figure~\ref{fig:fig_dens_xy_domain_series_fullrange} shows the stellar density in the orbital plane. After the binary hardens, the star is displaced but remains largely intact. Most of the stellar material remains gravitationally bound, with only $\simeq 1\%$ of the initial stellar mass leaving the computational domain. 

Figure \ref{fig:fig_temp_xy_domain_series} shows the stellar thermal response. The incoming black hole produces irreversible heating near the stellar surface during entry, whereas the later inspiral drives largely reversible compressional heating through the interior. We compute the net change using the mass-weighted temperature (specific internal energy),
\begin{equation}
 \langle T\rangle_M =
 \frac{\int \rho T\,dV}{\int \rho\,dV}, \label{eq:mass_weighted_temp}
\end{equation}
which rises by $27\%$ over the simulated evolution. Our simulation does not follow the much slower stellar response that comes afterward. Expansion and thermal relaxation occur on longer stellar-evolution timescales and will require a different treatment. We leave this for future work.

\subsection{Other systems}\label{sec:R3}
Run R2 begins from the same position as R1 but with a larger incoming speed. The broad outcome does not change. Over $\mathcal{O}(10$--$100)$ orbits, the intruding black hole loses enough orbital energy to sink to the stellar center and form a hard binary with the other one. At the end of the resolved evolution, the gravitational-wave-only coalescence estimate is $\mathcal{O}(10^3$--$10^4)\,\mathrm{yr}$.

\begin{figure}[!htb]
    \includegraphics[width=\columnwidth]{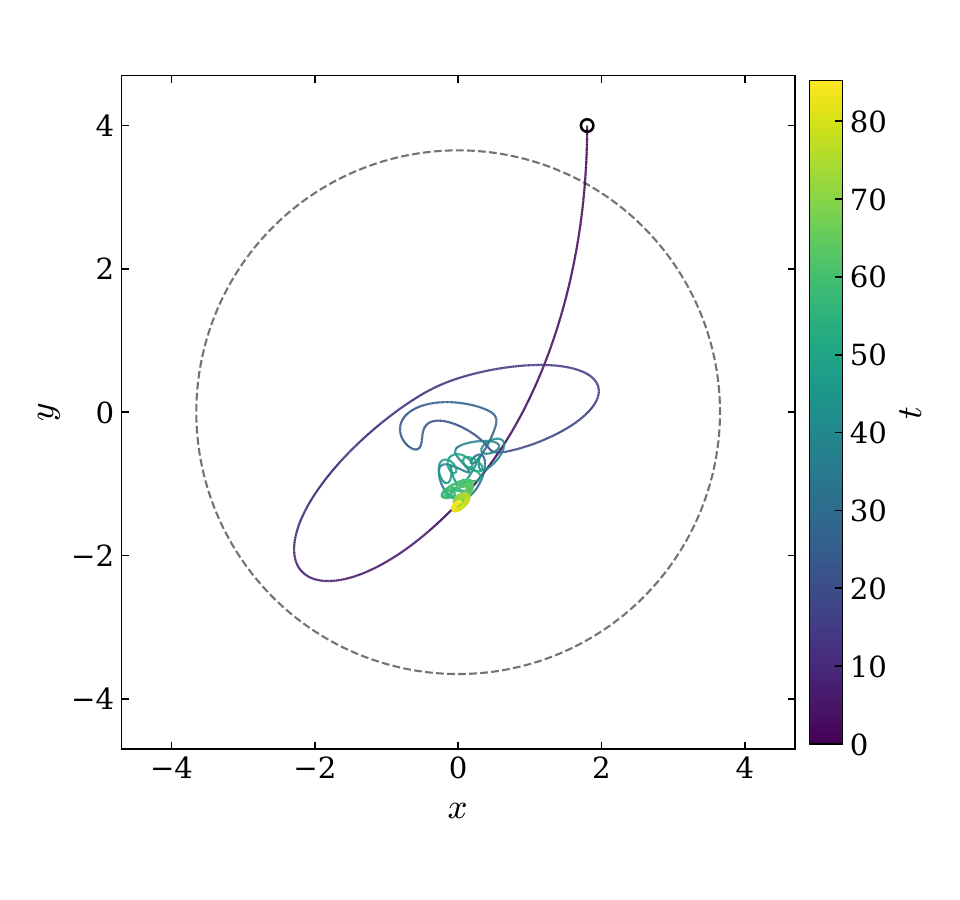}
    \caption{Orbital-plane trajectory of the intruding black hole in run R3,
    colored by time from release (open circle). The gray dashed circle marks the initial stellar surface. All quantities are in code units.}
    \label{fig:fig_traj_r1}
\end{figure}

\begin{figure}[!htb]
    \includegraphics[width=\columnwidth]{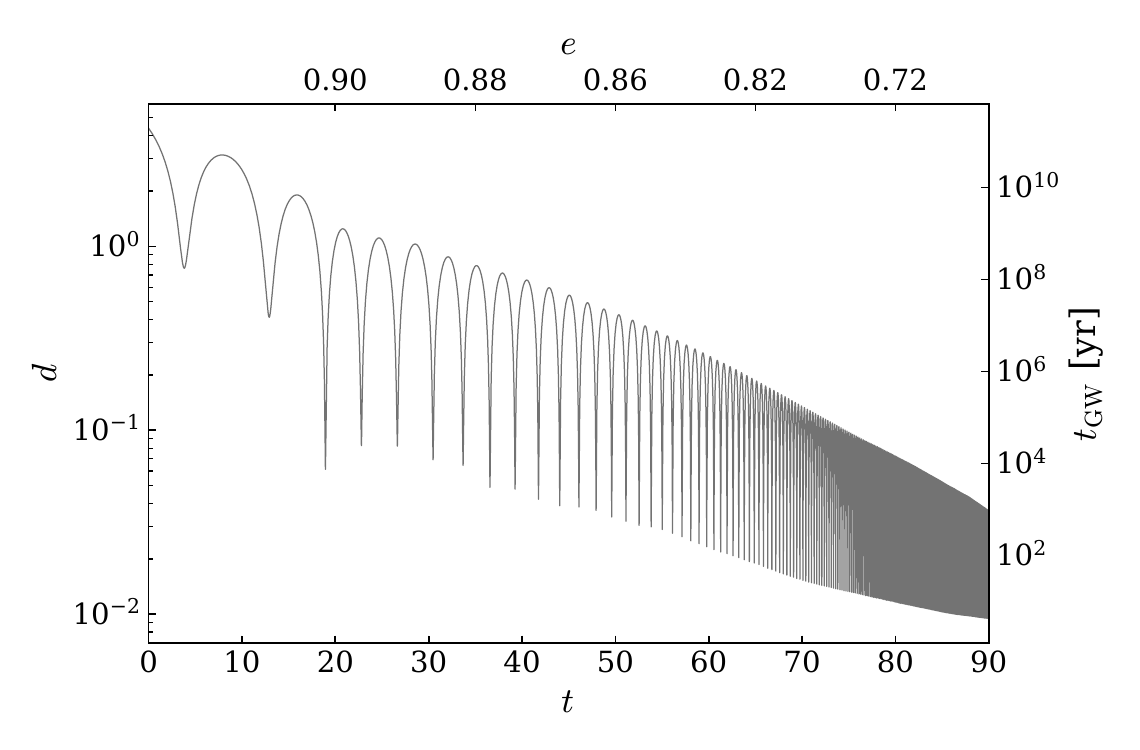}
    \caption{Similar to Fig.~\ref{fig:sep_res3}, orbital decay in run R3. The binary retains a nonnegligible eccentricity at the end of the simulation.}
    \label{fig:fig_sep_r1}
\end{figure}

R3 is captured as well, but its orbital history is more eccentric. Figures~\ref{fig:fig_traj_r1} and \ref{fig:fig_sep_r1} show rapid decay over $\mathcal{O}(100)$ cycles, again driven by dynamical friction. The inferred gravitational-wave coalescence time decreases to $\mathcal{O}(10^2$--$10^3)\,\mathrm{yr}$ by the end of the run. Unlike R1, the orbital eccentricity remains substantial, reaching $e\simeq0.67$ at the final time. This system therefore supports the findings of \citet{Hu:2026ica}.

\begin{figure}[!htb]
    \includegraphics[width=\columnwidth]{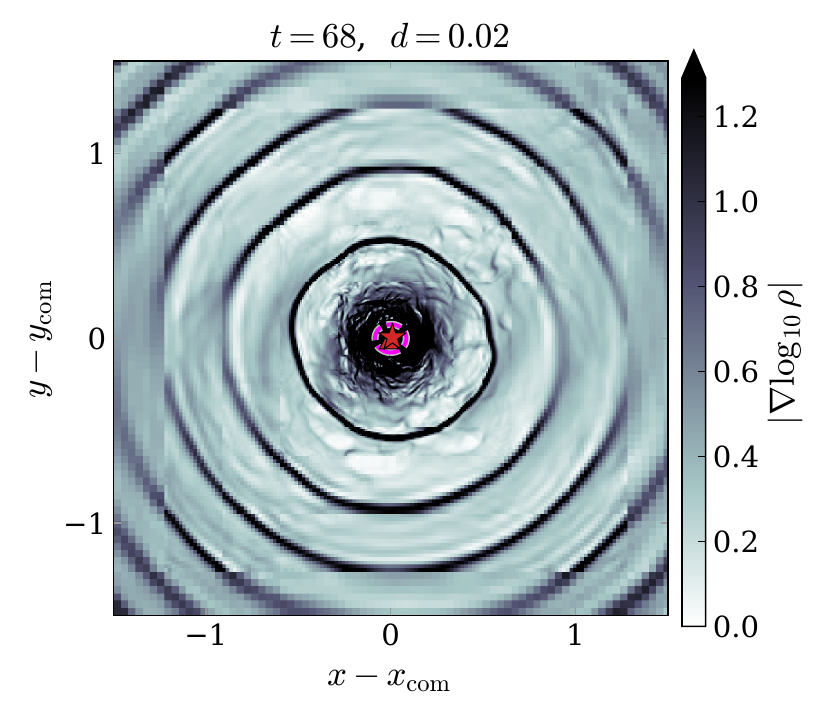}
    \caption{Similar to Fig.~\ref{fig:spiral_late}, but for run R3 at the single late time $t=68$. The red star and triangle mark the two black holes. The wave field is dominated by the $m=0$ radial oscillation of the star. All quantities are in code units.}
    \label{fig:fig_spiral_late_r1}
\end{figure}

The star itself again survives the encounter with little mass loss: only $0.92\%$ leaves the system, while the mass-weighted temperature rises by $12\%$. The stellar response, however, has a different morphology. Figure~\ref{fig:fig_spiral_late_r1} is dominated by an $m=0$ response rather than the $m=2$ spiral seen in R1 (Figure~\ref{fig:spiral_late}). We attribute this to the faster, more strongly deflected passage of the black hole through the stellar interior, which excites a radial oscillation of the star and produces the observed pattern.

\begin{figure*}[!htb]
    \includegraphics[width=0.98\textwidth]{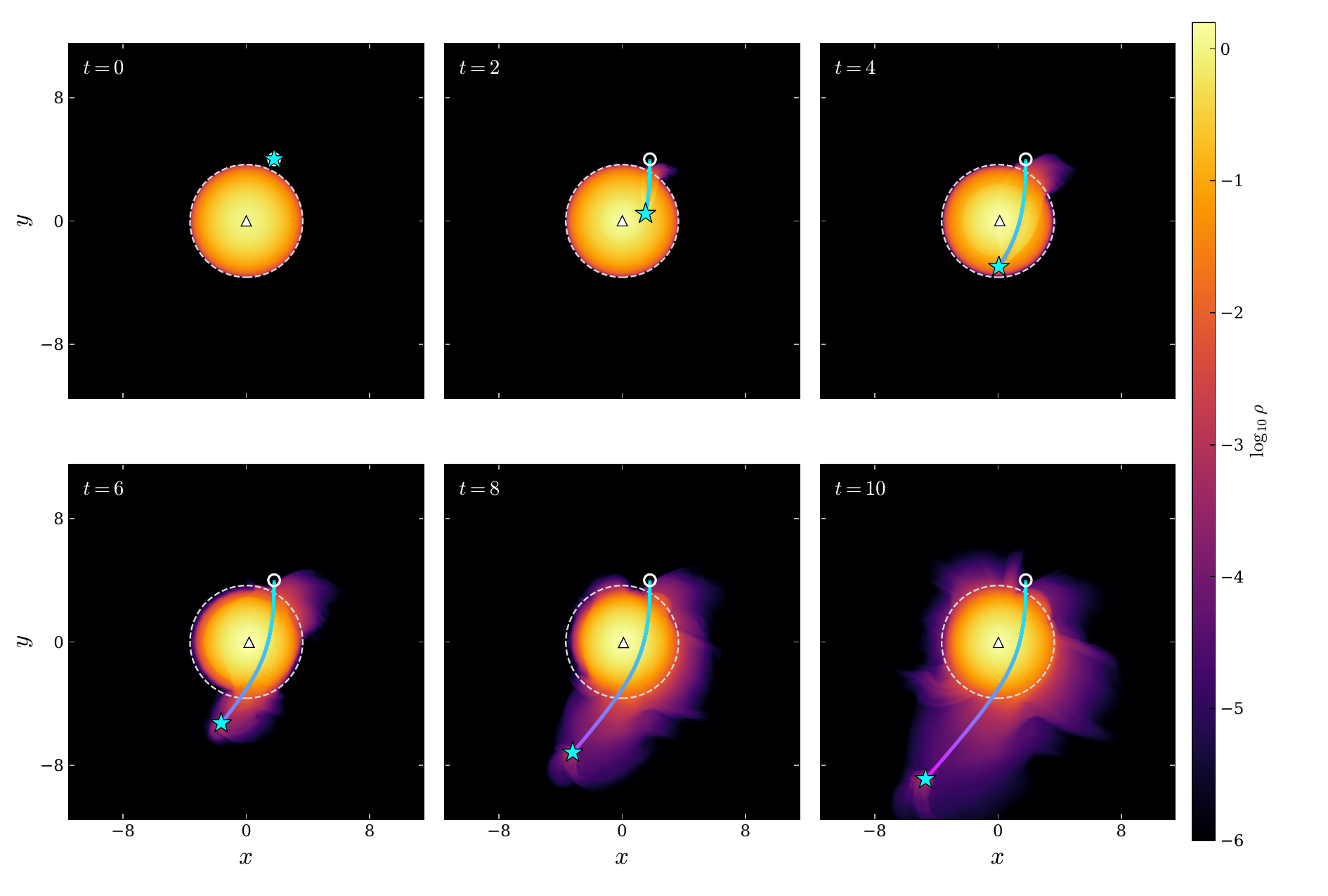}
    \caption{Stellar density profile for run R4, in which the intruding black hole enters the star at high velocity and subsequently escapes. The cyan curve traces the trajectory of the intruding black hole. The inner white triangle marks the other black hole, which remains nearly unperturbed throughout the encounter.}
    \label{fig:fig_traj_r2}
\end{figure*}

Finally, R4 provides a contrasting high-velocity case. It starts from the same position as R3, but the exterior black hole enters faster. As shown in Fig.~\ref{fig:fig_traj_r2}, the black hole passes through the star and escapes the system. Dynamical friction is not strong enough to capture the intruder. This flyby gives no appreciable kick to the central black hole and only weakly disturbs the stellar structure.

\section{Summary and Discussion}\label{sec:conclusion}
We have investigated encounters between stellar-mass black holes and massive AGN stars using three-dimensional hydrodynamic simulations. Across a range of encounter configurations, dynamical friction efficiently removes orbital angular momentum, causing the intruding black hole to sink toward the stellar center on the stellar dynamical timescale 
$\mathcal{O}(10^4)$s. 
Over this short timescale, the amount of gas accreted onto the 
black holes remains negligible compared with their masses. The later-arriving
black hole can then form a bound binary with a black hole already 
embedded in the star. The post-capture orbital evolution depends on the initial encounter geometry and velocity: some binaries circularize rapidly, whereas others retain substantial eccentricity. For a sufficiently high-velocity encounter, however, the intruding black hole can pass through the star and escape without being captured.

Once a bound binary has formed, further orbital decay can bring the system to merger. If the subsequent evolution is driven solely by gravitational-wave emission, the binaries produced in our simulations have coalescence timescales of $\lesssim 10^{3-4}$ yr. 
Such binaries can therefore subsequently evolve into sources detectable by gravitational-wave observatories \citep{Hu:2026ica, Cantiello2026}.
Despite differences in numerical methods and stellar modeling, our AthenaK simulations yield inspiral timescales broadly consistent with previous GIZMO simulations employing both MESA stellar profiles and different polytropic equations of state \citep{Shi:2026kci,Hu:2026ica}.
This general agreement supports the robustness of this binary-formation mechanism.

We also find that the rapid inspiral does not catastrophically disrupt the host star. The star remains gravitationally bound and loses only $\lesssim\mathcal{O}(1\%)$ of its mass. Its interior is compressed and heated during the encounter, raising the mass-weighted temperature by $\sim10$--$30\%$. Although this temperature temporarily enhances the nuclear 
burning rate through the CNO cycle, the elevated radiation pressure 
subsequently causes the expansion of the core, leading to adiabatic cooling, a reduction in the nuclear energy-generation rate, and an eventual return toward thermal equilibrium \citep{JiangInPrep}.

The late-time orbital dynamics in our current simulations are contaminated by the adopted sink and gravitational-softening parameters, preventing us from reliably following the binaries into the gravitational-wave-dominated regime and allowing us to place only an upper limit on the binary coalescence time.
A natural next step is therefore to perform higher-resolution, zoom-in simulations of the black hole binary and its immediate environment using more accurate treatments of accretion and small-scale gravitational dynamics. Our present simulations provide self-consistent initial conditions for such follow-up studies. Ultimately, fully general-relativistic simulations will be required to follow the merger within the host star and determine how the coalescence and gravitational-wave emission affect the stellar structure and whether it produces an observable electromagnetic counterpart. Such calculations could help characterize the multimessenger signatures of these black hole mergers.

Beyond these numerical extensions, another intriguing direction is to investigate repeated black hole captures and the resulting hierarchical merger process.
Such a pathway is possible because the recoil velocities of merger remnants are generally smaller than the escape velocity from the stellar core   \citep{Gonzalez2007,Baker:2006vn,Lousto:2007db,Baker:2007gi}, allowing them to remain gravitationally bound even if they are kicked into the envelope \citep{Hu:2026ica}. Since the estimated upper limit on the black hole merger timescale $\lesssim 10^{3-4}\,\mathrm{yr}$ is generally shorter than the encounter timescale between black holes and massive stars across most regions of AGN disks \citep{ChenLin2024}, a surviving main-sequence star may typically host at most one black hole merger remnant at a given time. Repeated encounters could nevertheless lead to successive generations of mergers. It will therefore be particularly interesting to determine how many generations of hierarchical mergers can occur before the host star either loses its ability to capture additional black holes or the growing embedded black hole becomes sufficiently massive to disrupt the star.

\section{Acknowledgements}
We thank Yanlong Shi and Qingru Hu for sharing the results of their simulations in advance of 
their publication.  We also thank Honghuan Jiang and Zhenghou Xu for useful conversations.
Research at Perimeter Institute is supported in part by the Government of Canada through the Department of Innovation, Science and Economic Development and by the Province of Ontario through the Ministry of Colleges, Universities, Research Excellence and Security.

\bibliography{references,journals}
\bibliographystyle{aasjournal}

\end{document}